\documentclass[final,3p,times]{elsarticle}
\usepackage{amsmath}
\usepackage{hyperref}
\usepackage{xcolor}

\newcommand{\sech}{\operatorname{sech}}

\biboptions{sort&compress}
\journal{Annals of Physics}

\hypersetup{colorlinks=true,linkcolor=blue,citecolor=blue,filecolor=blue,urlcolor=blue,breaklinks=true}

\begin{document}

\begin{frontmatter}

\title{Projected DKP spin-1 sectors in a $(2+1)$-dimensional static wormhole:
Self-duality, helicity content, and spectral correspondence}

\author[mymainaddress,mysecondaryaddress]{Luis B. Castro\corref{mycorrespondingauthor}}
\cortext[mycorrespondingauthor]{Corresponding author}
\ead{lrb.castro@ufma.br}

\address[mymainaddress]{Programa de P\'{o}s-graduação em F\'{i}sica, Universidade Federal do Maranh\~{a}o, Campus Universit\'{a}rio do Bacanga, 65080-805, S\~{a}o Lu\'{\i}s, Maranh\~{a}o, Brazil.}
\address[mysecondaryaddress]{Coordenaç\~{a}o do Curso de F\'{i}sica - Bacharelado, Universidade Federal do Maranh\~{a}o, Campus Universit\'{a}rio do Bacanga, 65080-805, S\~{a}o Lu\'{\i}s, Maranh\~{a}o, Brazil.}

\begin{abstract}

We use the projected spin-1 Duffin--Kemmer--Petiau (DKP) equation in
$2+1$ dimensions to clarify the physical content of the irreducible
$3\times3$ representations. The projected antisymmetric tensor is
algebraically dual to the projected vector, and the two twin representations
select conjugate self-dual covariant-curl sectors, which reduce to opposite
massive helicities in the flat-space limit. We then apply this structure to
the static wormhole problem previously studied with a single irreducible
representation. The first-order radial DKP system contains an exceptional
normalizable bound-state branch at $\omega^2=M^2$ that is absent from the
master-variable reduction, whereas the generic master and circular sectors
are related by a Darboux transformation. The latter preserves the transmission
and reflection probabilities, while the quasinormal modes are organized into
two pole families in the complex radial-momentum plane, leading to the
physical pairing $(\omega,-\omega^*)$ rather than a global $\pm\omega$
prescription.

\end{abstract}


\end{frontmatter}


\section{Introduction}

The Duffin--Kemmer--Petiau (DKP) formalism provides a first-order relativistic
wave equation for bosons of spin zero and spin one
\cite{Petiau1936,PR54:1114:1938,Kemmer1938,Kemmer1939}. In curved spacetime,
the physical content of the theory is most transparently expressed through its
projected components. In Riemannian geometry the minimally coupled spin-0 and
spin-1 sectors reproduce the Klein--Gordon and Proca dynamics, respectively,
whereas more general geometries such as Riemann--Cartan spacetime can modify
this correspondence~\cite{LUNARDI2001,GRG34:1941:2002}. In particular, the
projected spin-1 field obeys the curved-space Proca equation, in which the
Ricci tensor appears explicitly. The projected formulation is therefore the
natural starting point for identifying the physical vector degrees of freedom
and separating genuinely vectorial curvature effects from the scalar sector
\cite{EPJC86:902:2026}.

In $2+1$ dimensions the vector sector has an additional structure: an
antisymmetric rank-two tensor is Hodge-dual to a vector, and a first-order
self-dual massive-vector equation carries a single propagating excitation.
This is the setting of the self-dual model~\cite{PLB136:38:1984} and is
closely related to topologically massive gauge theory~\cite{AP140:372:1982}. The
DKP algebra likewise admits two twin irreducible $3\times3$ representations
\cite{CORSON1953}. Reference~\cite{EPL145:50002:2024} developed the
irreducible $3\times3$ DKP construction, including the adapted projector and
its relation to the reducible $6\times6$ formulation. What was not identified
there is the covariant-curl self-duality carried by each twin representation,
the corresponding helicity interpretation in the flat-space limit, and the
way in which the two sectors combine into a parity-even massive vector field.
These are the representation-theoretic issues addressed here in curved
spacetime.

A concrete test of these questions is provided by the static wormhole problem
of Ref.~\cite{AP477:169995:2025}. That work used one irreducible $3\times3$
DKP representation to study transmission and quasinormal modes of a massive
vector field and reduced the first-order system to a master equation for the
radial component $\Phi_0$. This construction raises two related questions.
First, does a single irreducible representation describe the full parity-even
massive vector field or only one self-dual sector? Second, does the reduction
to the master variable preserve the complete spectral content of the original
first-order DKP system? Answering these questions is necessary both for the
physical interpretation of the observables obtained in Ref.~\cite{AP477:169995:2025}
and for identifying any spectral information that may be lost when the
first-order system is reduced.

We address these questions using the projected DKP formulation. We construct
the two irreducible $3\times3$ representations in a common notation, identify
the projected vector and antisymmetric components, and show that their
algebraic duality becomes a covariant-curl self-duality equation. The two
self-dual sectors are then combined through the on-shell Proca decomposition,
which makes their parity relation and degree-of-freedom content explicit. We
subsequently return to the wormhole background and connect the full first-order
DKP system, the projected Proca equation, and the master and circular radial
descriptions through a first-order Darboux map.

We show that each irreducible $3\times3$ DKP representation corresponds to a
definite self-dual covariant-curl sector, while the parity-even massive vector
field requires the conjugate pair. Revisiting the wormhole problem of
Ref.~\cite{AP477:169995:2025}, we find that the first-order DKP system contains
an exceptional normalizable bound-state branch that is not captured by the
master-variable reduction, whereas the generic master and circular sectors are
related by a Darboux transformation and are almost isospectral. The scattering
probabilities remain unchanged between the paired sectors, while the
quasinormal-mode analysis requires the two pole families to be treated
separately in the complex momentum plane, leading to the physical pairing
$(\omega,-\omega^*)$.

The paper is organized as follows. In Sec.~\ref{sec:projected_dkp} we review
the projected DKP spin-1 sector and its relation to the curved-space Proca
equation. Section~\ref{sec:irreducible_representation} introduces the twin
irreducible $3\times3$ representations and derives the algebraic duality of
their projected components. In Sec.~\ref{sec:self_duality} this relation is
converted into the covariant self-dual equation and the parity-doublet
structure is established. Section~\ref{sec:darboux} applies the construction to
the static wormhole: the exceptional and generic sectors are separated first,
the Darboux factorization is then obtained, and its consequences for
scattering and quasinormal modes are analyzed. Finally,
Sec.~\ref{sec:conclusions} summarizes the results.

\section{Projected DKP spin-1 sector}
\label{sec:projected_dkp}

The first-order DKP equation in curved spacetime is \cite{GRG34:491:2002,GRG34:1941:2002,EPJC75:287:2015,EPJC76:61:2016,EPJC84:536:2024} ($\hbar=c=1$),
\begin{equation}
 \left(i \beta^\mu \nabla_\mu - M\right)\Psi=0,
 \label{eq:dkp}
\end{equation}
where $\beta^\mu=e^\mu{}_{\bar a}\beta^{\bar a}$ are the curved-space DKP matrices, which satisfy the algebra
\begin{equation}
 \beta^\mu\beta^\nu\beta^\lambda+\beta^\lambda\beta^\nu\beta^\mu
 =g^{\mu\nu}\beta^\lambda+g^{\lambda\nu}\beta^\mu.
 \label{eq:dkp_algebra}
\end{equation}
In $3+1$ dimensions, this algebra generates a set of 126 independent matrices. Its irreducible representations are the trivial one-dimensional representation, a five-dimensional representation describing spin-0 particles, and a ten-dimensional representation associated with spin-1 particles~\cite{UMEZAWA1956}.

In Eq.~\eqref{eq:dkp}, $M$ is the boson mass, and
\begin{equation}
 \nabla_\mu=\partial_\mu-\Gamma_\mu
\end{equation}
is the DKP covariant derivative. The covariant derivative acts on the DKP spinor and contains the spinorial
connection appropriate to the DKP algebra. This connection ensures the covariance of the DKP equation under local Lorentz transformations and is built from the spin connection of the tetrad frame. The spinorial connection is defined by
\begin{equation}
\Gamma_{\mu}=\frac{1}{2}\omega_{\mu\bar{a}\bar{b}}[\beta^{\bar{a}},\beta^{\bar{b}}],
\label{eq:spinorial_connection}
\end{equation}
where the spin connection coefficients are given by
\begin{equation}
\omega_{\mu}{}^{\bar{a}\bar{b}}=e_\alpha{}^{\bar a}e^\nu{}^{\bar{b}}\Gamma^{\alpha}_{\mu\nu}-e^\nu{}^{\bar{b}}\partial_{\mu}e_\nu{}^{\bar{a}},
\label{eq:spin_connection_coefficients}
\end{equation} 
with $\omega_{\mu}{}^{\bar{a}\bar{b}}=-\omega_{\mu}{}^{\bar{b}\bar{a}}$ and the Christoffel symbols are given by
\begin{equation}
\Gamma_{\mu\nu}^{\alpha}=\frac{g^{\alpha\lambda}}{2}\left(\partial_{\mu}g_{\lambda\nu}+\partial_{\nu}g_{\lambda\mu}-\partial_{\lambda}g_{\mu\nu}\right).
\label{eq:christoffel_symbols}
\end{equation}
The tetrads $e_\mu{}^{\bar a}$ satisfy
\begin{eqnarray}
\eta^{\bar{a}\bar{b}} &=& e_\mu{}^{\bar a}e_\nu{}^{\bar b}g^{\mu\nu},\label{eq:tetrad_relation_1}\\
g_{\mu\nu} &=& e_\mu{}^{\bar a}e_\nu{}^{\bar b}\eta_{\bar{a}\bar{b}},\label{eq:tetrad_relation_2}
\end{eqnarray}
and
\begin{equation}
e_\mu{}^{\bar a}e^\nu{}_{\bar a}=\delta_\mu^\nu, \qquad
e^\mu{}_{\bar a}e_\mu{}^{\bar b}=\delta_{\bar a}^{\bar b}.
\label{eq:tetrad_inverse_relations}
\end{equation}
These relations fix the conversion between coordinate and local Lorentz indices. Greek indices denote spacetime coordinates, whereas barred Latin indices label the local Lorentz frame and are not DKP-spinor indices. They are raised and lowered with $g_{\mu\nu}$ and $\eta_{\bar a\bar b}$, respectively.

In the projected formulation, the spin-1 sector is isolated by means of Umezawa's operators~\cite{UMEZAWA1956}. In four spacetime dimensions, their local-frame form is defined by
\begin{align}
R^{\bar a}
&=(\beta^{\bar 1})^2(\beta^{\bar 2})^2(\beta^{\bar 3})^2
\left(\beta^{\bar a}\beta^{\bar 0}-\eta^{\bar a\bar 0}\right),
\label{eq:umezawa_Ra}
\\
R^{\bar a\bar b}
&=R^{\bar a}\beta^{\bar b},
\label{eq:umezawa_Rab}
\end{align}
and their curved-space counterparts are
\begin{align}
R^\mu
&=e^\mu{}_{\bar a}R^{\bar a},
\label{eq:umezawa_Rmu}
\\
R^{\mu\nu}
&=e^\mu{}_{\bar a}e^\nu{}_{\bar b}R^{\bar a\bar b}.
\label{eq:umezawa_Rmunu}
\end{align}
These operators satisfy
\begin{align}
R^{\mu\nu} &=-R^{\nu\mu},
\label{eq:umezawa_antisymmetry}
\\
R^\mu S^{\nu\alpha}
&=g^{\mu\nu}R^\alpha-g^{\mu\alpha}R^\nu,
\label{eq:umezawa_vector}
\\
R^{\mu\nu}S^{\alpha\rho}
&=g^{\mu\rho}R^{\nu\alpha}
-g^{\mu\alpha}R^{\nu\rho}
+g^{\nu\alpha}R^{\mu\rho}
-g^{\nu\rho}R^{\mu\alpha},
\label{eq:umezawa_tensor}
\end{align}
where $S^{\mu\nu}=[\beta^\mu,\beta^\nu]$. These identities ensure that $R^\mu\Psi$ transforms as a vector and $R^{\mu\nu}\Psi$ as an antisymmetric rank-two tensor under local Lorentz transformations. We therefore denote the projected vector by $W^\mu=R^\mu\Psi$, while $R^{\mu\nu}\Psi$ denotes the corresponding projected antisymmetric component. Applying these operators to the DKP equation gives the projected relations
\begin{equation}
 R^\nu\Psi=\frac{i}{M}\nabla_\mu\left(R^{\nu\mu}\Psi\right),
 \label{eq:Rnu_relation}
\end{equation}
\begin{equation}
 R^{\nu\mu}\Psi=\frac{i}{M}U^{\mu\nu},
 \qquad
 U^{\mu\nu}=\nabla^\mu W^\nu-\nabla^\nu W^\mu.
 \label{eq:U_def}
\end{equation}
Since $U^{\mu\nu}$ is constructed entirely from $W^\mu$, Eq.~\eqref{eq:U_def} shows that the projected antisymmetric tensor is determined by the projected vector field and is therefore not an independent field. Substituting Eq.~\eqref{eq:U_def} into Eq.~\eqref{eq:Rnu_relation} gives the curved-space Proca equation
\begin{equation}
 \nabla_\mu U^{\mu\nu}+M^2 W^\nu=0.
 \label{eq:proca_first}
\end{equation}
We adopt the curvature convention $[\nabla_\mu,\nabla_\nu]W^\rho=\mathcal{R}^\rho{}_{\sigma\mu\nu}W^\sigma$, where $\mathcal{R}^\rho{}_{\sigma\mu\nu}$ is the Riemann curvature tensor~\cite{EPJC86:902:2026}. Commuting the covariant derivatives in Eq.~\eqref{eq:proca_first} then gives
\begin{equation}
 \left(\nabla_\alpha\nabla^\alpha+M^2\right)W^\nu
 -\mathcal{R}_\lambda{}^{\nu}W^\lambda
 -\nabla^\nu\left(\nabla_\mu W^\mu\right)=0.
 \label{eq:proca_pre_sub}
\end{equation}
Here $\mathcal{R}_\lambda{}^{\nu}=g^{\nu\rho}\mathcal{R}_{\lambda\rho}$ is the mixed Ricci tensor, with $\mathcal{R}_{\lambda\rho}=\mathcal{R}^{\mu}{}_{\lambda\mu\rho}$.
The last term in Eq.~\eqref{eq:proca_pre_sub} is fixed by the Proca equation itself. Taking the divergence of Eq.~\eqref{eq:proca_first} and using the antisymmetry of $U^{\mu\nu}$ gives $\nabla_\nu\nabla_\mu U^{\mu\nu}=\tfrac12[\nabla_\nu,\nabla_\mu]U^{\mu\nu}=0$, because the commutator reduces to contractions of the symmetric Ricci tensor with the antisymmetric tensor $U^{\mu\nu}$. Hence, for $M\neq0$,
\begin{equation}
 \nabla_\mu W^\mu=0,
 \label{eq:subsidiary}
\end{equation}
and so Eq.~\eqref{eq:proca_pre_sub} reduces to
\begin{equation}
  \left(\nabla_\alpha\nabla^\alpha+M^2\right)W^\nu
 -\mathcal{R}_\lambda{}^{\nu}W^\lambda=0 .
  \label{eq:proca_projected}
\end{equation}
This is the projected Proca equation obtained from the spin-1 DKP sector~\cite{EPJC86:902:2026}. It is important that the Ricci tensor appears explicitly in the spin-1 sector. This term is absent in the corresponding minimally coupled scalar equation and is a purely vectorial curvature contribution. The form \eqref{eq:proca_projected} is the equation to be compared with any second-order equation
obtained directly from a particular DKP representation. In particular, it makes
explicit the genuinely vectorial part of the dynamics, namely the curvature coupling. In the flat-space limit, $\mathcal{R}_\lambda{}^\nu=0$ and $\nabla_\mu\to\partial_\mu$, so Eq.~\eqref{eq:proca_projected} reduces to the usual massive-vector equation $(\Box+M^2)W^\nu=0$, together with $\partial_\mu W^\mu=0$.

\section{Irreducible \(3\times3\) DKP representation in \(2+1\) dimensions}
\label{sec:irreducible_representation}

The standard four-dimensional Umezawa construction reviewed in Sec.~\ref{sec:projected_dkp} contains the factor $(\beta^{\bar3})^2$. In $(2+1)$ dimensions there is no $\beta^{\bar3}$, and the projector must therefore be adapted to the irreducible three-dimensional representation. The spin-1 sector of the $(2+1)$-dimensional DKP algebra admits a twin pair of irreducible $3\times3$ representations~\cite{CORSON1953,EPL145:50002:2024}. It is convenient to treat them simultaneously through
\begin{equation}
\beta^{\bar0}_{(\sigma)}=
\begin{pmatrix}
0&0&0\\
0&0&i\sigma\\
0&-i\sigma&0
\end{pmatrix},
\qquad
\beta^{\bar1}_{(\sigma)}=
\begin{pmatrix}
0&0&-1\\
0&0&0\\
1&0&0
\end{pmatrix},
\qquad
\beta^{\bar2}_{(\sigma)}=
\begin{pmatrix}
0&1&0\\
-1&0&0\\
0&0&0
\end{pmatrix},
\label{eq:beta_sigma_family}
\end{equation}
with $\sigma=\pm1$. The choice $\sigma=+1$ reproduces the representation used in Ref.~\cite{AP477:169995:2025} and in Ref.~\cite{EPL145:50002:2024}; $\sigma=-1$ gives its twin representation. For the algebraic discussion below we write the three-component DKP spinor as
\begin{equation}
\psi=
\begin{pmatrix}
\Phi_0\\
\Phi_+\\
\Phi_-
\end{pmatrix}.
\label{eq:psi3}
\end{equation}
Here $\psi$ denotes only the three-component object on which the local matrices in Eq.~\eqref{eq:beta_sigma_family} act.

For the family \eqref{eq:beta_sigma_family}, the $(2+1)$-dimensional counterpart of the standard Umezawa factor is the idempotent operator
\begin{equation}
\Pi=(\beta^{\bar1}_{(\sigma)})^2(\beta^{\bar2}_{(\sigma)})^2
=
\begin{pmatrix}
1&0&0\\
0&0&0\\
0&0&0
\end{pmatrix}.
\label{eq:Pi_2p1}
\end{equation}
This adapted projector, which satisfies $\Pi^2=\Pi$, is the one used in the irreducible $3\times3$ construction of Ref.~\cite{EPL145:50002:2024}. Using this idempotent factor, the corresponding $(2+1)$-dimensional Umezawa vector operators are defined by
\begin{equation}
R^{\bar a}=\Pi\left(\beta^{\bar a}_{(\sigma)}\beta^{\bar0}_{(\sigma)}-\eta^{\bar a\bar0}\right),
\qquad \bar a=0,1,2,
\label{eq:R_def_2p1}
\end{equation}
with $\eta_{\bar a\bar b}=\mathrm{diag}(+,-,-)$. Acting with these operators on the spinor $\psi$ in Eq.~\eqref{eq:psi3} gives
\begin{equation}
R^{\bar0}\psi=
\begin{pmatrix}-\Phi_0\\0\\0\end{pmatrix},
\qquad
R^{\bar1}\psi=
\begin{pmatrix}i\sigma\Phi_+\\0\\0\end{pmatrix},
\qquad
R^{\bar2}\psi=
\begin{pmatrix}i\sigma\Phi_-\\0\\0\end{pmatrix}.
\label{eq:Rpsi_sigma}
\end{equation}
Since Sec.~\ref{sec:projected_dkp} established that $R^{\bar a}\psi$ transforms as a local Lorentz vector, Eq.~\eqref{eq:Rpsi_sigma} identifies its three projected components explicitly. Their nonzero entries are fixed exactly, including the representation-dependent phases, by the choice of $\sigma$.

The corresponding antisymmetric sector is obtained from
\begin{equation}
R^{\bar a\bar b}=R^{\bar a}\beta^{\bar b}_{(\sigma)} .
\label{eq:Rab_def_2p1}
\end{equation}
For the same three-component spinor one obtains
\begin{equation}
R^{\bar0\bar1}\psi=
\begin{pmatrix}\Phi_-\\0\\0\end{pmatrix},
\qquad
R^{\bar0\bar2}\psi=
\begin{pmatrix}-\Phi_+\\0\\0\end{pmatrix},
\qquad
R^{\bar1\bar2}\psi=
\begin{pmatrix}-i\sigma\Phi_0\\0\\0\end{pmatrix}.
\label{eq:Rabpsi_sigma}
\end{equation}
With the orientation convention $\epsilon^{\bar0\bar1\bar2}=+1$, Eqs.~\eqref{eq:Rpsi_sigma} and \eqref{eq:Rabpsi_sigma} give the algebraic identity
\begin{equation}
R^{\bar a\bar b}\psi
=\sigma i\,\epsilon^{\bar a\bar b}{}_{\bar c}R^{\bar c}\psi .
\label{eq:algebraic_duality}
\end{equation}
Thus, once the orientation and the definition \eqref{eq:Rab_def_2p1} are fixed, the sign $\sigma$ is fixed by the irreducible representation; it is not an additional dynamical choice. In particular, the representation used in Ref.~\cite{AP477:169995:2025} corresponds to $\sigma=+1$.

The two $3\times3$ reductions of the reducible $6\times6$ formulation were referred to as equivalent in Ref.~\cite{EPL145:50002:2024}. In the stricter fixed-generator similarity sense, however, the two members of Eq.~\eqref{eq:beta_sigma_family} are inequivalent. Indeed,
\begin{equation}
\operatorname{tr}\!\left(
\beta^{\bar0}_{(\sigma)}
\beta^{\bar1}_{(\sigma)}
\beta^{\bar2}_{(\sigma)}
\right)=i\sigma ,
\label{eq:twin_trace}
\end{equation}
and this trace is invariant under similarity transformations while changing sign between $\sigma=+1$ and $\sigma=-1$. The two representations are therefore naturally paired as twins, but they are distinguished algebraically by the opposite sign in Eq.~\eqref{eq:algebraic_duality}.

\section{Self-duality and helicity content}
\label{sec:self_duality}

The algebraic identity \eqref{eq:algebraic_duality} can now be converted into a spacetime relation. The local alternating symbol of Sec.~\ref{sec:irreducible_representation} defines the curved-space Levi-Civita tensor through
\begin{equation}
\epsilon^{\mu\nu\rho}
=e^\mu{}_{\bar a}e^\nu{}_{\bar b}e^\rho{}_{\bar c}
\epsilon^{\bar a\bar b\bar c},
\qquad
\nabla_\lambda\epsilon^{\mu\nu\rho}=0.
\label{eq:levi_civita_curved}
\end{equation}
Thus, \(\epsilon^{\mu\nu\rho}\) is a spacetime tensor, distinct from the local-frame symbol \(\epsilon^{\bar a\bar b\bar c}\). Combining Eq.~\eqref{eq:algebraic_duality} with the projected DKP relation \eqref{eq:U_def} gives
\begin{equation}
U^{\mu\nu}=-\sigma M\epsilon^{\mu\nu}{}_{\rho}W^\rho.
\label{eq:U_self_dual}
\end{equation}
Equation~\eqref{eq:U_self_dual} is the spacetime form of the algebraic duality established in Sec.~\ref{sec:irreducible_representation}. It shows that, within each irreducible \(3\times3\) representation, the antisymmetric field strength is algebraically determined by the projected vector field. To express this relation as a first-order equation for \(W^\mu\), we take its Hodge dual.

In $(2+1)$ dimensions the Hodge dual of an antisymmetric rank-two tensor is a vector. For the field strength \(U^{\mu\nu}\), we define
\begin{equation}
(\star U)^\mu
\equiv\frac{1}{2}\epsilon^\mu{}_{\alpha\beta}U^{\alpha\beta}
=\frac{1}{2}\epsilon^\mu{}_{\alpha\beta}
\left(\nabla^\alpha W^\beta-\nabla^\beta W^\alpha\right).
\label{eq:hodge_dual_field_strength}
\end{equation}
Interchanging the dummy indices \(\alpha\) and \(\beta\) in the second term and using the antisymmetry of the Levi-Civita tensor gives
\begin{equation}
(\star U)^\mu
=\epsilon^\mu{}_{\alpha\beta}\nabla^\alpha W^\beta.
\label{eq:hodge_dual_curl}
\end{equation}
To express Eq.~\eqref{eq:hodge_dual_curl} in a compact form, we define the covariant-curl operator by
\begin{equation}
(\mathcal{C}W)^\mu
\equiv\epsilon^\mu{}_{\alpha\beta}\nabla^\alpha W^\beta.
\label{eq:covariant_curl_definition}
\end{equation}
With this definition, Eq.~\eqref{eq:hodge_dual_curl} simply reads \((\star U)^\mu=(\mathcal{C}W)^\mu\).

On the other hand, taking the Hodge dual of Eq.~\eqref{eq:U_self_dual} gives
\begin{equation}
(\star U)^\mu
=-\frac{\sigma M}{2}
\epsilon^\mu{}_{\alpha\beta}
\epsilon^{\alpha\beta}{}_{\rho}W^\rho.
\label{eq:hodge_dual_self_dual_relation}
\end{equation}
Using the three-dimensional contraction identity \(\frac{1}{2}\epsilon^\mu{}_{\alpha\beta}\epsilon^{\alpha\beta}{}_{\rho}=\delta^\mu{}_{\rho}\), together with Eqs.~\eqref{eq:hodge_dual_curl} and \eqref{eq:covariant_curl_definition}, one obtains
\begin{equation}
(\mathcal{C}W)^\mu
=\epsilon^\mu{}_{\alpha\beta}\nabla^\alpha W^\beta
=-\sigma M W^\mu.
\label{eq:self_dual_equation}
\end{equation}
Equation~\eqref{eq:self_dual_equation} is the first-order massive self-dual equation of the type introduced in Ref.~\cite{PLB136:38:1984}. The self-dual model is closely related to, but is not the same formulation as, the topologically massive gauge theory of Ref.~\cite{AP140:372:1982}; both descriptions carry the same single massive spin-1 excitation.

Equation~\eqref{eq:self_dual_equation} also makes the meaning of the helicity label precise. In flat $(2+1)$-dimensional spacetime the two eigenvalues of \(\mathcal{C}\) correspond to the two signs of the massive spin, or helicity, associated with the $SO(2)$ little group. In a general curved background there is no global Poincar\'e little group, so the rigorous label is the sign of the covariant-curl eigenvalue. We use the term ``helicity sector'' in this sense; it reduces to the standard massive $(2+1)$-dimensional helicity in the flat-space limit. With the conventions adopted here, the representation label and the curl eigenvalue have opposite signs: \(\sigma=+1\) gives \(\mathcal{C}W=-MW\), whereas \(\sigma=-1\) gives \(\mathcal{C}W=+MW\).

The self-dual equation is more restrictive than the second-order Proca system. Taking the divergence of Eq.~\eqref{eq:self_dual_equation} gives \(\nabla_\mu W^\mu=0\) for \(M\neq0\). Indeed, using Eq.~\eqref{eq:levi_civita_curved}, the divergence of the covariant curl can be written as \(\nabla_\mu(\mathcal{C}W)^\mu=\frac{1}{2}\epsilon^{\mu\alpha\beta}[\nabla_\mu,\nabla_\alpha]W_\beta=0\), where the last equality follows from the first Bianchi identity. Moreover, using the curvature convention fixed in Sec.~\ref{sec:projected_dkp}, the square of the covariant curl satisfies
\begin{equation}
(\mathcal{C}^2W)^\mu
=-\nabla_\alpha\nabla^\alpha W^\mu
+\nabla^\mu\!\left(\nabla_\alpha W^\alpha\right)
+\mathcal{R}_\lambda{}^\mu W^\lambda.
\label{eq:curl_squared_identity}
\end{equation}
Applying \(\mathcal{C}\) once more to Eq.~\eqref{eq:self_dual_equation} yields
\begin{equation}
(\mathcal{C}^2W)^\mu=M^2W^\mu.
\label{eq:curl_squared_self_dual}
\end{equation}
Together with \(\nabla_\mu W^\mu=0\), Eqs.~\eqref{eq:curl_squared_identity} and \eqref{eq:curl_squared_self_dual} give
\begin{equation}
\left(\nabla_\alpha\nabla^\alpha+M^2\right)W^\mu
-\mathcal{R}_\lambda{}^\mu W^\lambda=0,
\label{eq:self_dual_implies_proca}
\end{equation}
which is precisely the projected Proca equation \eqref{eq:proca_projected}. Thus, each self-dual sector is a restricted solution sector of the projected Proca theory.

Conversely, let \(W^\mu\) satisfy the projected Proca equation together with the subsidiary condition \eqref{eq:subsidiary}. Equation~\eqref{eq:curl_squared_identity} then gives \(\mathcal{C}^2W=M^2W\). On the Proca solution space one may therefore define
\begin{equation}
\mathcal{P}_\pm
=\frac{1}{2}\left(1\pm\frac{\mathcal{C}}{M}\right).
\label{eq:curl_projectors}
\end{equation}
These operators satisfy, on shell,
\begin{align}
\mathcal{P}_\pm^2&=\mathcal{P}_\pm,
\label{eq:curl_projectors_idempotent}
\\
\mathcal{P}_+\mathcal{P}_-&=0,
\qquad
\mathcal{P}_++\mathcal{P}_-=1.
\label{eq:curl_projectors_complete}
\end{align}
Defining
\begin{equation}
W^\mu_{(\pm)}=(\mathcal{P}_\pm W)^\mu,
\label{eq:self_dual_components}
\end{equation}
one obtains
\begin{equation}
W^\mu=W^\mu_{(+)}+W^\mu_{(-)},
\qquad
\mathcal{C}W_{(\pm)}=\pm M W_{(\pm)}.
\label{eq:proca_self_dual_decomposition}
\end{equation}
Hence the projected Proca field decomposes covariantly into the two self-dual sectors. Comparing Eq.~\eqref{eq:proca_self_dual_decomposition} with Eq.~\eqref{eq:self_dual_equation}, the \(\sigma=+1\) DKP representation realizes the negative-curl sector \(W_{(-)}\), while \(\sigma=-1\) realizes the positive-curl sector \(W_{(+)}\).

An orientation-reversing parity transformation changes the sign of the Levi-Civita tensor and therefore sends \(\mathcal{C}\to-\mathcal{C}\). It consequently interchanges \(W_{(+)}\) and \(W_{(-)}\). A single self-dual sector is not parity-even by itself, whereas the pair is closed under parity. In terms of the twin irreducible DKP representations of Sec.~\ref{sec:irreducible_representation},
\begin{equation}
\text{parity-even massive vector field in }(2+1)
=(3\times3)_{\sigma=+1}\oplus(3\times3)_{\sigma=-1}.
\label{eq:parity_doublet}
\end{equation}
This decomposition also reproduces the degree-of-freedom counting. A massive Proca field in $(2+1)$ dimensions has three vector components subject to one subsidiary condition and therefore two physical degrees of freedom. Each first-order self-dual sector carries one of them. Consequently, the $3\times3$ representation used in Ref.~\cite{AP477:169995:2025} is not wrong; it is an irreducible self-dual representation. However, by itself it does not represent the full parity-even massive vector field.

\section{Darboux factorization and spectral data}
\label{sec:darboux}

\subsection{Radial DKP system and exceptional sector}
\label{sec:radial_dkp_exceptional}

We now apply the projected formulation to the static wormhole background of Ref.~\cite{AP477:169995:2025},
\begin{equation}
 ds^2=dt^2-d\nu^2-f^2(\nu)d\varphi^2,
 \qquad
 f(\nu)=b\cosh(\nu/a),
 \label{eq:wh_metric}
\end{equation}
where $a$ is the curvature radius and $b$ is the throat radius. We use the orthonormal coframe
\begin{equation}
 e^{\hat 0}=dt,
 \qquad
 e^{\hat\nu}=d\nu,
 \qquad
 e^{\hat\varphi}=f(\nu)d\varphi,
 \label{eq:wormhole_frame}
\end{equation}
so that hatted indices refer to this adapted local frame. The wormhole manifold admits a global orthonormal frame, and since $f(\nu)=b\cosh(\nu/a)>0$ for all $\nu$, the coframe chosen above is nonsingular throughout the geometry. With the curvature convention fixed in Sec.~\ref{sec:projected_dkp}, the nonvanishing mixed Ricci components are
\begin{equation}
 \mathcal{R}_{\hat\nu}{}^{\hat\nu}
 =\mathcal{R}_{\hat\varphi}{}^{\hat\varphi}
 =\frac{1}{a^2},
 \qquad
 \mathcal{R}_{\hat0}{}^{\hat0}=0.
 \label{eq:wormhole_ricci}
\end{equation}
For either member of the twin family in Eq.~\eqref{eq:beta_sigma_family}, we take the separated field
\begin{equation}
 \Psi_{(\sigma)}(t,\nu,\varphi)=
 \frac{e^{-i\omega t}e^{i\ell\varphi}}{\sqrt{f(\nu)}}
 \begin{pmatrix}
 \Phi_0(\nu)\\
 \Phi_+(\nu)\\
 \Phi_-(\nu)
 \end{pmatrix},
 \qquad \ell\in\mathbb Z.
 \label{eq:ansatz}
\end{equation}
Using Eq.~\eqref{eq:Rpsi_sigma}, the local components of the projected vector are
\begin{equation}
 W^{\hat0}=-\frac{e^{-i\omega t+i\ell\varphi}}{\sqrt f}\,\Phi_0,
 \qquad
 W^{\hat\nu}=\frac{i\sigma e^{-i\omega t+i\ell\varphi}}{\sqrt f}\,\Phi_+,
 \qquad
 W^{\hat\varphi}=\frac{i\sigma e^{-i\omega t+i\ell\varphi}}{\sqrt f}\,\Phi_-.
 \label{eq:projected_components_wormhole}
\end{equation}

Reference~\cite{AP477:169995:2025} derives the radial DKP system using the representation that corresponds here to $\sigma=+1$. In the twin family \eqref{eq:beta_sigma_family}, only $\beta^{\bar0}$ carries the factor $\sigma$, while the spatial matrices are unchanged. For the static coframe \eqref{eq:wormhole_frame}, the spin-connection contribution is built from the spatial commutator and is therefore also independent of $\sigma$. With the time dependence in Eq.~\eqref{eq:ansatz}, the twin system is consequently obtained from the $\sigma=+1$ equations by the replacement $\omega\to\sigma\omega$. The complete family of radial equations is therefore
\begin{subequations}
\label{eq:first_order_full}
\begin{align}
 M\Phi_0&=-i\widetilde D\Phi_- -L\Phi_+,
 \label{eq:first_order_full_a}
 \\
 M\Phi_+&=i\sigma\omega\Phi_-+L\Phi_0,
 \label{eq:first_order_full_b}
 \\
 M\Phi_-&=-i\sigma\omega\Phi_+ +iD\Phi_0,
 \label{eq:first_order_full_c}
\end{align}
\end{subequations}
where, for compactness,
\begin{equation}
 D=\frac{d}{d\nu}-\frac{f'}{2f},
 \qquad
 \widetilde D=\frac{d}{d\nu}+\frac{f'}{2f},
 \qquad
 L=\frac{\ell}{f}.
 \label{eq:radial_operators}
\end{equation}
The last two equations in \eqref{eq:first_order_full} constitute a $2\times2$ linear system for the spatial components $\Phi_+$ and $\Phi_-$. The reconstruction of these components from $\Phi_0$ fails precisely when the coefficient matrix becomes singular, namely when
\begin{equation}
 \det
 \begin{pmatrix}
 M & -i\sigma\omega\\
 i\sigma\omega & M
 \end{pmatrix}
 =M^2-\omega^2=0.
 \label{eq:radial_subsystem_determinant}
\end{equation}
Thus the generic reconstruction applies for $\omega^2\neq M^2$, whereas the exceptional case $\omega^2=M^2$ must be analyzed directly in the first-order system.

We first consider this exceptional case. Writing $\sigma\omega=sM$, with $s=\pm1$, the two equations that normally reconstruct the spatial components become linearly dependent. Their compatibility requires
\begin{equation}
 (D-sL)\Phi_0=0.
 \label{eq:exceptional_phi0_condition}
\end{equation}
The corresponding nonvanishing solution is
\begin{equation}
 \Phi_0(\nu)=C\sqrt{f(\nu)}
 \exp\!\left[
 \frac{s\ell a}{b}\operatorname{gd}\!\left(\frac{\nu}{a}\right)
 \right],
 \label{eq:exceptional_phi0_solution}
\end{equation}
where $\operatorname{gd}(x)=\int_0^x\sech y\,dy$ is the Gudermannian. Since $\operatorname{gd}(x)\to\pm\pi/2$ as $x\to\pm\infty$, whereas $\sqrt f\sim e^{|\nu|/(2a)}$, the branch in Eq.~\eqref{eq:exceptional_phi0_solution} is not normalizable. The normalizable branch therefore has $\Phi_0=0$. In this branch, Eqs.~\eqref{eq:first_order_full_b} and \eqref{eq:first_order_full_c} give $\Phi_+=is\Phi_-$, while Eq.~\eqref{eq:first_order_full_a} reduces to
\begin{equation}
 (\widetilde D+sL)\Phi_-=0.
 \label{eq:exceptional_phim_equation}
\end{equation}
The normalizable solution is therefore
\begin{equation}
 \Phi_0^{(0)}=0,
 \qquad
 \Phi_+^{(0)}=is\Phi_-^{(0)},
 \qquad
 \Phi_-^{(0)}(\nu)=\mathcal N f^{-1/2}(\nu)
 \exp\!\left[
 -\frac{s\ell a}{b}\operatorname{gd}\!\left(\frac{\nu}{a}\right)
 \right].
 \label{eq:exceptional_dkp_mode}
\end{equation}
Because $f^{-1/2}\sim e^{-|\nu|/(2a)}$, this state is square integrable for every $\ell$ and both values of $s$. Substitution of Eq.~\eqref{eq:exceptional_dkp_mode} into the subsidiary condition \eqref{eq:subsidiary} reduces it precisely to Eq.~\eqref{eq:exceptional_phim_equation}. Its physical character can also be checked with the DKP current $J^\mu=\tfrac12\bar\Psi\beta^\mu\Psi$, where $\bar\Psi=\Psi^\dagger\eta^0$ and $\eta^0=2(\beta^{\bar0})^2-1$. For the state \eqref{eq:exceptional_dkp_mode},
\begin{equation}
 J^\nu=0,
 \qquad
 J^0=\frac{\sigma s}{f}\left|\Phi_-^{(0)}\right|^2.
 \label{eq:exceptional_current}
\end{equation}
The corresponding charge $Q=2\pi\sigma s\int_{-\infty}^{+\infty}|\Phi_-^{(0)}(\nu)|^2d\nu$ is finite and nonzero. Together with square integrability, the vanishing radial current identifies Eq.~\eqref{eq:exceptional_dkp_mode} as a genuine DKP bound state. Since $\Phi_0^{(0)}=0$, this branch is not contained in the master equation used for the generic reconstruction in Ref.~\cite{AP477:169995:2025}. That work identified the fundamental $(n,\ell)=(0,0)$ QNM frequency $\omega=M$; the first-order analysis above reveals, in addition, a normalizable bound-state branch at $\omega^2=M^2$ for every $\ell$.

\subsection{Generic sector and Darboux factorization}
\label{sec:generic_darboux}

For $\omega^2\neq M^2$, the determinant \eqref{eq:radial_subsystem_determinant} is nonzero and Eqs.~\eqref{eq:first_order_full_b} and \eqref{eq:first_order_full_c} can be inverted. One obtains
\begin{equation}
 \Phi_+=\frac{\sigma\omega D\Phi_0-ML\Phi_0}{\omega^2-M^2},
 \qquad
 \Phi_-=\frac{i\left(\sigma\omega L\Phi_0-MD\Phi_0\right)}{\omega^2-M^2}.
 \label{eq:phi_pm_recon}
\end{equation}
The reconstruction formulas simplify when the two spatial components are combined in a circular basis. Indeed, adding the first expression in Eq.~\eqref{eq:phi_pm_recon} to $is$ times the second gives
\begin{equation}
 \Phi_+ +is\Phi_-
 =\frac{\sigma\omega+sM}{\omega^2-M^2}(D-sL)\Phi_0.
 \label{eq:circular_combination_from_reconstruction}
\end{equation}
We therefore define $\chi_s\equiv\Phi_+ +is\Phi_-$, with $s=\pm1$. Since $\omega^2-M^2=(\sigma\omega-sM)(\sigma\omega+sM)$, Eq.~\eqref{eq:circular_combination_from_reconstruction} becomes
\begin{equation}
 \chi_s=\frac{1}{\sigma\omega-sM}
 (D-sL)\Phi_0.
 \label{eq:darboux_relation}
\end{equation}
Here $s$ labels the two circular components and is distinct from the representation/self-duality label $\sigma$ introduced in Secs.~\ref{sec:irreducible_representation} and \ref{sec:self_duality}.

The corresponding circular components of the projected vector field are
\begin{equation}
 W_s=\frac{1}{\sqrt2}\left(W^{\hat\nu}+isW^{\hat\varphi}\right),
 \label{eq:Ws_def}
\end{equation}
with $s=\pm1$. Using the projected components in Eq.~\eqref{eq:projected_components_wormhole} together with the definition of $\chi_s$, one obtains
\begin{equation}
 W_s=\frac{i\sigma}{\sqrt{2f}}
 e^{-i\omega t+i\ell\varphi}\chi_s.
 \label{eq:Ws_chi_relation}
\end{equation}
Writing
\begin{equation}
 W_s=e^{-i\omega t+i\ell\varphi}w_s(\nu),
 \qquad
 w_s=\frac{u_s}{\sqrt f},
 \label{eq:us_definition}
\end{equation}
gives
\begin{equation}
 u_s=\frac{i\sigma}{\sqrt2}\chi_s.
 \label{eq:us_chi_relation}
\end{equation}
Combining Eqs.~\eqref{eq:darboux_relation} and \eqref{eq:us_chi_relation} therefore gives the direct first-order relation between the master variable and the circular radial variable,
\begin{equation}
 u_s=\frac{i\sigma}{\sqrt2\,(\sigma\omega-sM)}
 (D-sL)\Phi_0.
 \label{eq:us_darboux_map}
\end{equation}
Thus the circular DKP combination and the circular component of the projected vector describe the same radial degree of freedom up to the constant factor displayed above.

Having established the first-order relation between $\Phi_0$ and $u_s$, we now determine the second-order equations intertwined by this map. Eliminating $\Phi_\pm$ from the first-order system \eqref{eq:first_order_full} gives the master equation for $\Phi_0$. The same equation follows independently from the time component of the projected Proca equation \eqref{eq:proca_projected}. Since $\mathcal R_{\hat\lambda}{}^{\hat0}=0$, Eq.~\eqref{eq:projected_components_wormhole} reduces the time component to
\begin{equation}
 \left[
 \frac{d^2}{d\nu^2}
 +\frac{f'}{f}\frac{d}{d\nu}
 +\omega^2-M^2-\frac{\ell^2}{f^2}
 \right]
 \left(\frac{\Phi_0}{\sqrt f}\right)=0.
 \label{eq:proca_time_radial}
\end{equation}
For $f=b\cosh(\nu/a)$ this becomes
\begin{equation}
 \Phi_0''+
 \left[
 \kappa^2-
 \left(\frac{1}{4a^2}+\frac{\ell^2}{b^2}\right)
 \sech^2(\nu/a)
 \right]\Phi_0=0,
 \label{eq:phi0_master}
\end{equation}
Equation~\eqref{eq:phi0_master} is precisely the master equation obtained from the irreducible $3\times3$ DKP system in Ref.~\cite{AP477:169995:2025}. Its independent derivation here from the temporal component of the projected Proca equation provides a direct consistency check between the first-order DKP and projected second-order descriptions. Here
\begin{equation}
 \kappa^2=\omega^2-M^2-\frac{1}{4a^2}.
 \label{eq:kappa}
\end{equation}

The spatial components of Eq.~\eqref{eq:proca_projected} mix $W^{\hat\nu}$ and $W^{\hat\varphi}$. The circular basis \eqref{eq:Ws_def} diagonalizes this mixing and gives
\begin{equation}
 w_s''+\frac{f'}{f}w_s'
 +\left[
 \omega^2-M^2-\frac{(\ell+s f')^2}{f^2}
 +\frac{f''}{f}
 \right]w_s=0.
 \label{eq:ws_first_derivative}
\end{equation}
Equation~\eqref{eq:ws_first_derivative} has the same general structure as the circular-component equation obtained in Ref.~\cite{EPJC86:902:2026} for the projected spin-1 field, here restricted to $2+1$ dimensions. To remove the first-derivative term and expose the effective radial potential, we use the rescaling introduced in Eq.~\eqref{eq:us_definition}. With $w_s=u_s/\sqrt f$ and $f=b\cosh(\nu/a)$, Eq.~\eqref{eq:ws_first_derivative} becomes
\begin{equation}
 u_s''+
 \left[
 \kappa^2+
 \left(\frac{3}{4a^2}-\frac{\ell^2}{b^2}\right)\sech^2(\nu/a)
 -\frac{2s\ell}{ab}\sech(\nu/a)\tanh(\nu/a)
 \right]u_s=0.
 \label{eq:us_partner}
\end{equation}
Equation~\eqref{eq:us_partner} is the Schr\"odinger-like equation for the circular radial variable $u_s$. The two circular channels differ through the term proportional to $s\ell\,\sech(\nu/a)\tanh(\nu/a)$, while the remaining potential terms are common.

The first-order map~\eqref{eq:us_darboux_map} already contains the differential combination $D-sL$, and the same structure organizes the two second-order radial equations. To make this factorization explicit, we define
\begin{equation}
 A_s=\frac{d}{d\nu}-\mathcal Q_s,
 \qquad
 A_s^\dagger=-\frac{d}{d\nu}-\mathcal Q_s,
 \label{eq:darboux_operators}
\end{equation}
where $\mathcal Q_s=f'/(2f)+s\ell/f$. Hence $A_s=D-sL$, while $A_s^\dagger$ is the formal adjoint with respect to the $d\nu$ measure. These operators satisfy
\begin{equation}
 A_s^\dagger A_s
 =-\frac{d^2}{d\nu^2}+\mathcal Q_s^2+\mathcal Q_s',
 \qquad
 A_sA_s^\dagger
 =-\frac{d^2}{d\nu^2}+\mathcal Q_s^2-\mathcal Q_s'.
 \label{eq:darboux_products}
\end{equation}
Substituting $f=b\cosh(\nu/a)$ into Eq.~\eqref{eq:darboux_products} reproduces the effective potentials in Eqs.~\eqref{eq:phi0_master} and \eqref{eq:us_partner}, respectively. The two radial equations can therefore be written in the factorized form
\begin{align}
 A_s^\dagger A_s\Phi_0&=(\omega^2-M^2)\Phi_0,
 \label{eq:factor_master}
 \\
 A_sA_s^\dagger u_s&=(\omega^2-M^2)u_s.
 \label{eq:factor_partner}
\end{align}
The generic relation \eqref{eq:us_darboux_map} is the corresponding first-order Darboux map~\cite{PR251:267:1995}.

The factorized form also identifies the exceptional state obtained independently from the first-order DKP system in Sec.~\ref{sec:radial_dkp_exceptional}. At $\omega^2=M^2$, the nonnormalizable branch in Eq.~\eqref{eq:exceptional_phi0_solution} satisfies $A_s\Phi_0=0$, whereas the physical exceptional mode \eqref{eq:exceptional_dkp_mode} gives a normalizable circular state satisfying $A_s^\dagger u_s^{(0)}=0$. Thus the normalizable zero mode of the partner operator is not an artifact of the second-order factorization; it is precisely the DKP bound state obtained directly from the first-order system. The two one-dimensional radial problems are therefore almost isospectral: their generic spectra are paired, while the circular partner contains the additional normalizable bound state at $\omega^2=M^2$ for every $\ell$.

The reflection $\varphi\to-\varphi$ is an isometry of the static wormhole background defined by the metric~\eqref{eq:wh_metric}. Under this transformation, the angular factor $e^{i\ell\varphi}$ gives $\ell\to-\ell$, while the circular components are exchanged, $W_s\leftrightarrow W_{-s}$, so that $s\to-s$. Since the reflection reverses the orientation of the spatial section, the covariant curl changes sign and the two self-dual representations are exchanged, $\sigma\to-\sigma$. Consistently, the master equation~\eqref{eq:phi0_master} for $\Phi_0$ is insensitive to these sign changes: the representation label enters the reduction through $(\sigma\omega)^2=\omega^2$, while the circular and angular labels enter quadratically through $(s\ell)^2=\ell^2$, with $\sigma^2=s^2=1$. The circular partner equation~\eqref{eq:us_partner} is likewise invariant because its sign-sensitive term depends only on the product $s\ell$. Therefore the reflection realizes $(\sigma,\ell,s)\to(-\sigma,-\ell,-s)$, showing explicitly the geometric origin of the degeneracy between the twin self-dual sectors in this background.

\subsection{Scattering data}
\label{sec:scattering_data}

Scattering states require oscillatory solutions in the two asymptotic regions of the wormhole. We therefore begin with the asymptotic form of the master equation~\eqref{eq:phi0_master} for $\Phi_0$. Since $\sech(\nu/a)\to0$ as $|\nu|\to\infty$, Eq.~\eqref{eq:phi0_master} reduces to $\Phi_0''+\kappa^2\Phi_0\simeq0$. Hence the parameter $\kappa$ defined in Eq.~\eqref{eq:kappa} acquires the physical meaning of the asymptotic radial momentum. Propagating scattering states require real $\kappa$, or equivalently $\kappa^2>0$. It is therefore convenient to define the propagation threshold by
\begin{equation}
 \omega_0^2=M^2+\frac{1}{4a^2},
 \label{eq:omega0_threshold}
\end{equation}
so that $\kappa^2=\omega^2-\omega_0^2$ and the scattering continuum corresponds to $\omega^2>\omega_0^2$.

The shift $1/(4a^2)$ has a geometric origin in the radial reduction. In passing from Eq.~\eqref{eq:proca_time_radial} to the master equation~\eqref{eq:phi0_master}, the $f^{-1/2}$ rescaling generates a constant term $K/4=-1/(4a^2)$ for the wormhole spatial curvature $K=-f''/f=-1/a^2$, equivalently the positive threshold gap $-K/4=1/(4a^2)$ appearing in Eq.~\eqref{eq:omega0_threshold}. This shift is therefore not the explicit vectorial Ricci coupling emphasized in Sec.~\ref{sec:projected_dkp}.

To compare the master and circular scattering channels, it remains to verify that the same asymptotic momentum governs both. In the circular equation~\eqref{eq:ws_first_derivative}, one has $f''/f\to1/a^2$, $\ell/f\to0$, and $f'/f\to\pm1/a$ as $|\nu|\to\infty$. Consequently, the asymptotic $f''/f$ contribution is cancelled by the $(f'/f)^2$ term contained in $(\ell+s f')^2/f^2$. After the rescaling $w_s=u_s/\sqrt f$ introduced in Eq.~\eqref{eq:us_definition}, the circular equation reduces to $u_s''+\kappa^2u_s\simeq0$. Thus the master and circular channels share the same asymptotic radial momentum $\kappa$.

Transmission and reflection probabilities are among the main observables analyzed in Ref.~\cite{AP477:169995:2025}. Since the present formulation reinterprets the irreducible $3\times3$ DKP system as a single self-dual sector and restores the parity-even vector field by pairing it with the conjugate sector, it is important to determine whether this refinement changes the scattering observables obtained from the master equation. Having established that the master and circular channels have the same asymptotic momentum, we can compare their scattering data. Normalize a wave incident from the left by
\begin{equation}
 \Phi_0(\nu\to-\infty)=e^{i\kappa\nu}+r_0e^{-i\kappa\nu},
 \qquad
 \Phi_0(\nu\to+\infty)=t_0e^{i\kappa\nu}.
 \label{eq:scattering_asymptotics}
\end{equation}
Because the asymptotic momentum is the same on both sides, the corresponding master-channel flux ratios are $T_{\Phi_0,\ell}=|t_0|^2$ and $R_{\Phi_0,\ell}=|r_0|^2$. For the wormhole profile,
\begin{equation}
 \frac{f'}{2f}\to-\frac{1}{2a}
 \quad(\nu\to-\infty),
 \qquad
 \frac{f'}{2f}\to+\frac{1}{2a}
 \quad(\nu\to+\infty),
 \qquad
 \frac{\ell}{f}\to0
 \quad(|\nu|\to\infty).
 \label{eq:darboux_asymptotics}
\end{equation}
These limits determine directly how the Darboux map acts on the asymptotic plane waves. From Eq.~\eqref{eq:darboux_operators},
\begin{equation}
 A_s\to\frac{d}{d\nu}+\frac{1}{2a}
 \quad(\nu\to-\infty),
 \qquad
 A_s\to\frac{d}{d\nu}-\frac{1}{2a}
 \quad(\nu\to+\infty).
 \label{eq:darboux_operator_asymptotics}
\end{equation}
Equation~\eqref{eq:us_darboux_map} has the form $u_s=C_sA_s\Phi_0$, where $C_s=i\sigma/[\sqrt2(\sigma\omega-sM)]$ is independent of $\nu$. At $\nu\to-\infty$, the incident and reflected waves are therefore multiplied by $i\kappa+1/(2a)$ and $-i\kappa+1/(2a)$, respectively, whereas at $\nu\to+\infty$ the transmitted wave is multiplied by $i\kappa-1/(2a)$. The common factor $C_s$ cancels when the incident circular amplitude is normalized to unity, giving
\begin{equation}
 t_s=\frac{i\kappa-\frac{1}{2a}}
 {i\kappa+\frac{1}{2a}}\,t_0,
 \qquad
 r_s=\frac{-i\kappa+\frac{1}{2a}}
 {i\kappa+\frac{1}{2a}}\,r_0.
 \label{eq:phase_factors}
\end{equation}
For real $\kappa$, both Darboux prefactors have unit modulus. Since the two asymptotic regions also have the same momentum, the corresponding flux ratios satisfy
\begin{equation}
 T_{s,\ell}(\omega)=T_{\Phi_0,\ell}(\omega),
 \qquad
 R_{s,\ell}(\omega)=R_{\Phi_0,\ell}(\omega).
 \label{eq:TR_equal}
\end{equation}
Once the Darboux relation has been established, this equality follows from the asymptotic properties of the intertwining operator: the regularity $f(\nu)\ge b>0$, the decay $\ell/f\to0$, the opposite limits of $f'/(2f)$ in Eq.~\eqref{eq:darboux_asymptotics}, and the common asymptotic momentum $\kappa$. It therefore does not require the two effective potentials to have the same detailed shape in the interior.

The only spectral mismatch between the Darboux partners found in Sec.~\ref{sec:generic_darboux} does not affect this scattering equivalence. The exceptional energy $\omega^2=M^2$ corresponds to $\kappa^2=-1/(4a^2)$ and therefore lies below the continuum threshold~\eqref{eq:omega0_threshold}; it is not part of the real-$\kappa$ scattering sector. Hence the equality in Eq.~\eqref{eq:TR_equal} applies to all propagating states.

The circular channels consequently inherit the transmission probability obtained from the master equation in Ref.~\cite{AP477:169995:2025},
\begin{equation}
 T_{s,\ell}(\omega)=
 \frac{
 \cosh\!\left(2\pi a\sqrt{\omega^2-\omega_0^2}\right)-1
 }{
 \cosh\!\left(2\pi a\sqrt{\omega^2-\omega_0^2}\right)
 +\cosh\!\left(2\pi a|\ell|/b\right)
 }.
 \label{eq:transmission_final}
\end{equation}
The absence of either $s$ or $\sigma$ in Eq.~\eqref{eq:transmission_final} makes the scattering degeneracy of the two circular and twin self-dual sectors explicit in this background.

\subsection{Quasinormal modes}
\label{sec:qnm}

Quasinormal modes characterize the resonant response of the open wormhole scattering problem and were one of the principal spectral quantities analyzed in Ref.~\cite{AP477:169995:2025}. Since the present formulation reinterprets the irreducible $3\times3$ DKP system as a single self-dual sector and relates the circular channel to the master equation through a Darboux transformation, it is necessary to determine whether the QNM spectrum obtained from the master variable is preserved and how the exceptional factorization sector enters the pole structure.

A QNM is selected by purely outgoing behavior at both asymptotic ends of the wormhole, equivalently by a pole of the analytically continued transmission amplitude. Unlike the real-energy probability in Eq.~\eqref{eq:transmission_final}, this resonance condition depends on the phase and sheet information of the complex amplitude. We therefore return to the master transmission amplitude and define
\begin{equation}
 \kappa(\omega)=\sqrt{\omega^2-\omega_0^2}
 \label{eq:kappa_sheet}
\end{equation}
on the physical positive-frequency scattering branch by $\kappa>0$ for real $\omega>\omega_0$. The outgoing solutions are then continued from this branch into the lower half of the $\kappa$ plane. With the time dependence $e^{-i\omega t}$ adopted in Eq.~\eqref{eq:ansatz}, the outgoing conditions are
\begin{equation}
 \Phi_0(\nu\to-\infty)\sim e^{-i\kappa\nu},
 \qquad
 \Phi_0(\nu\to+\infty)\sim e^{+i\kappa\nu}.
 \label{eq:qnm_outgoing}
\end{equation}
For real $\kappa$, the first wave propagates toward the left asymptotic region and the second toward the right; their analytic continuation defines the QNM boundary conditions for complex $\kappa$.

Imposing Eq.~\eqref{eq:qnm_outgoing} is equivalent to requiring the incoming-wave coefficient to vanish, or, equivalently, to locating the poles of the analytically continued master transmission amplitude. For the P\"oschl--Teller problem associated with Eq.~\eqref{eq:phi0_master}, the two pole families are
\begin{equation}
 \kappa_{n\ell}^{(\eta)}
 =\eta\frac{|\ell|}{b}
 -\frac{i}{a}\left(n+\frac12\right).
 \label{eq:kappa_qnm_poles}
\end{equation}
Here $\eta=\pm1$ labels the two pole families and $n=0,1,2,\ldots$.
The two families must be mapped separately to the frequency plane through $\omega^2=\omega_0^2+\kappa^2$; appending a global $\pm$ sign after using only one of the two $\kappa$ families loses the sheet information. It is convenient to define
\begin{equation}
 \Omega_{n\ell}=\sqrt{
 \omega_0^2+
 \left[
 \frac{|\ell|}{b}-\frac{i}{a}\left(n+\frac12\right)
 \right]^2}.
 \label{eq:Omega_qnm}
\end{equation}
The branch is chosen such that $\operatorname{Re}\Omega_{n\ell}\ge0$ and $\operatorname{Im}\Omega_{n\ell}\le0$.
The physical frequency families are therefore
\begin{equation}
 \omega_{n\ell}^{(+)}=\Omega_{n\ell},
 \qquad
 \omega_{n\ell}^{(-)}=-\Omega_{n\ell}^{*}.
 \label{eq:qnm_final}
\end{equation}
Both families satisfy $\operatorname{Im}\omega_{n\ell}^{(\pm)}\le0$.
Thus the spectral symmetry is $\omega\to-\omega^*$, rather than a literal global $\omega\to-\omega$. This corrects the undifferentiated global $\pm$ prescription used in Ref.~\cite{AP477:169995:2025}: with the convention $e^{-i\omega t}$, the literal negative of a damped complex frequency generally lies in the upper half of the $\omega$ plane, whereas the physical conjugate partner is $-\omega^*$. For $\ell=0$, the two $\kappa$ families in Eq.~\eqref{eq:kappa_qnm_poles} coalesce on the imaginary axis. If the radicand in Eq.~\eqref{eq:Omega_qnm} is negative, the two frequency families in Eq.~\eqref{eq:qnm_final} also coalesce into the single root with negative imaginary part; when it is nonnegative, they reduce to the two real roots related by $\omega\to-\omega$.

Having established the master-channel QNM poles, we can use the same Darboux relation employed in the scattering analysis to determine how these poles are represented in the circular channel and how the exceptional factorization state enters the analytic structure. The prefactor in Eq.~\eqref{eq:phase_factors} has a pole at $\kappa=+i/(2a)$ and a zero at $\kappa=-i/(2a)$. These are distinct points of the $\kappa$ plane although both satisfy $\omega^2=M^2$. The upper-half-plane pole is the bound-state pole associated with the normalizable DKP mode in Eq.~\eqref{eq:exceptional_dkp_mode}; as shown in Sec.~\ref{sec:radial_dkp_exceptional}, this state exists for every $\ell$ and is not a QNM.

The lower-half-plane zero becomes relevant for the exceptional QNM point. For $\ell=0$ and $n=0$, Eq.~\eqref{eq:kappa_qnm_poles} gives $\kappa=-i/(2a)$, precisely the zero of the Darboux prefactor. Reference~\cite{AP477:169995:2025} identified the corresponding frequency $\omega=M$ as the fundamental QNM. The present analysis shows that this point must be treated separately from the generic Darboux correspondence: it lies exactly at the factorization energy, where the reconstruction from the master variable degenerates. Moreover, because the two master pole families coalesce for $\ell=0$, the coincident $\Gamma$-function poles of the P\"oschl--Teller amplitude give a double master pole at $n=0$, while the simple Darboux zero reduces the circular amplitude to a simple pole at the same QNM location. Thus the frequency identified in Ref.~\cite{AP477:169995:2025} remains a QNM pole, but its analytic multiplicity and its relation to the full DKP system are exceptional rather than generic.

This exceptional QNM pole must also be distinguished from the genuine DKP bound state found in Sec.~\ref{sec:radial_dkp_exceptional}. The latter occurs at the distinct point $\kappa=+i/(2a)$ and exists for every $\ell$, although it has the same value $\omega^2=M^2$. For $\ell\neq0$, or for $n\ge1$, no QNM pole in Eq.~\eqref{eq:kappa_qnm_poles} coincides with the Darboux zero, and the master poles are inherited directly by the circular channel.

The spectral relation is therefore precise. The master and circular radial problems have identical scattering probabilities and the same QNM pole locations; the exceptional $(n,\ell)=(0,0)$ QNM has reduced pole multiplicity in the circular amplitude, while the circular partner additionally contains the normalizable bound state at $\omega^2=M^2$. The two twin representations carry parity-related copies of this structure. Their radial profiles and scattering phases can differ, but the transmission probabilities and QNM frequencies are degenerate in the present wormhole background.

\section{Conclusions}
\label{sec:conclusions}

We have analyzed the projected DKP spin-1 sector in a \((2+1)\)-dimensional static wormhole background, with emphasis on the physical content of the irreducible \(3\times3\) representations. The projected quantities \(R^\mu\psi\) form a vector field in the local frame, so the components \(\Phi_0\), \(\Phi_+\), and \(\Phi_-\) have a direct projected-vector interpretation rather than being arbitrary auxiliary amplitudes.

For each irreducible representation, the projected antisymmetric tensor is algebraically dual to the projected vector. This relation reduces the projected DKP equations to a first-order self-dual covariant-curl equation. A single \(3\times3\) representation therefore describes one self-dual covariant-curl sector, which reduces to one massive helicity sector in the flat-space limit, whereas the parity-even massive vector field requires the two conjugate self-dual sectors. Relative to the irreducible \(3\times3\) construction discussed in Ref.~\cite{EPL145:50002:2024}, the present analysis identifies this self-dual/helicity content explicitly and shows how the conjugate sectors are paired by parity; the wormhole problem then provides the curved-space setting in which their radial and spectral relation can be tested.

In the wormhole background, the first-order DKP system separates naturally into the exceptional sector \(\omega^2=M^2\) and the generic sector \(\omega^2\neq M^2\). The exceptional sector contains a genuine normalizable DKP bound state for every \(\ell\), with \(\Phi_0=0\), and is therefore absent from the master-equation description employed in Ref.~\cite{AP477:169995:2025}. In the generic sector, the master variable and the circular projected-vector component are related by a first-order Darboux map, and their second-order radial operators form an almost-isospectral pair. Thus the master equation correctly describes the generic part of a single irreducible self-dual sector, but it does not exhaust the full first-order DKP spectrum because it misses the exceptional bound-state branch.

The Darboux relation also determines which scattering and resonance results are preserved. For real asymptotic momentum, its asymptotic factors have unit modulus, so the circular and master channels have identical transmission and reflection probabilities even though their scattering phases can differ. The continuum threshold \(\omega_0^2=M^2+1/(4a^2)\) is the geometric/rescaling gap of the wormhole radial problem, not a signature of the explicit vector Ricci coupling. For quasinormal modes, the poles must first be organized as the two families in the complex \(\kappa\) plane and only then mapped to the frequency plane, giving the physical pairing \((\omega,-\omega^*)\) rather than a global \(\pm\omega\) prescription. This corrects the branch assignment used in Ref.~\cite{AP477:169995:2025}. The point \((n,\ell)=(0,0)\), for which \(\omega^2=M^2\), is also exceptional: the coincident master poles at \(\kappa=-i/(2a)\) have their multiplicity reduced by the Darboux zero in the circular amplitude, while the normalizable DKP bound state lies at the distinct point \(\kappa=+i/(2a)\) and exists for every \(\ell\).

The projected formulation therefore does more than clarify the representation-theoretic interpretation of the \(3\times3\) DKP sector: in this wormhole background it distinguishes the spectral information that is robust under the passage between the master and circular descriptions from the exceptional information that is lost in the master reduction employed in Ref.~\cite{AP477:169995:2025}. The two twin representations remain parity-related and spectrally degenerate here, so the parity-even massive vector field contains both self-dual sectors rather than a new set of generic scattering or QNM frequencies. At the same time, the exceptional \(\Phi_0=0\) bound-state branch shows that the full first-order system is essential when the Darboux reconstruction becomes singular. This separation between generic degeneracy and exceptional spectral content is specific to the structure established for the present background and provides the appropriate basis for extending the analysis to settings in which the conjugate sectors need not remain degenerate.

\section*{Acknowledgments}
The author is grateful to the anonymous Referees for their careful reading and constructive criticisms and suggestions, which significantly improved the clarity and presentation of the manuscript. 

\section*{Funding}
This work was supported in part by means of funds provided by CNPq, Brazil, Grant Nos. 308172\slash 2023-0 and CNPq\slash Universal\slash 420896/2025-2, FAPEMA and CAPES (Finance code 001).


\end{document}